\documentclass[a4paper]{spie}  

\usepackage{amsmath,amsfonts,amssymb}
\usepackage{graphicx}
\usepackage[colorlinks=true, allcolors=blue]{hyperref}

\title{RISTRETTO: high-resolution spectroscopy at the diffraction limit of the VLT}

\author[a]{Christophe Lovis}
\author[a]{Nicolas Blind}
\author[a]{Bruno Chazelas}
\author[a]{Jonas G. K\"uhn}
\author[a]{Ludovic Genolet}
\author[a]{Ian Hughes}
\author[a]{Micha\"el Sordet}
\author[a]{Robin Schnell}
\author[a,b]{Martin Turbet}
\author[c,d]{Thierry Fusco}
\author[c,d]{Jean-Fran\c cois Sauvage}
\author[a]{Maddalena Bugatti}
\author[a]{Nicolas Billot}
\author[a]{Janis Hagelberg}
\author[a]{Eddy Hocini}
\author[e,f,g,h]{Olivier Guyon}
\author[i]{Christoph Mordasini}

\affil[a]{D\'epartement d'Astronomie, Universit\'e de Gen\`eve, Chemin Pegasi 51, CH-1290 Versoix, Switzerland}
\affil[b]{Laboratoire de M\'et\'eorologie Dynamique, IPSL, CNRS, Sorbonne Universit\'e, 4 place Jussieu, F-75252 Paris Cedex 05, France}
\affil[c]{Laboratoire d'Astrophysique de Marseille, CNRS, CNES, Aix Marseille Universit\'e, 38 rue Fr\'ed\'eric Joliot Curie, F-13013 Marseille, France}
\affil[d]{DOTA, ONERA, Universit\'e Paris Saclay, F-91123 Palaiseau, France}
\affil[e]{Subaru Telescope, National Astronomical Observatory of Japan, National Institutes of Natural Sciences (NINS), 650 North Aohoku Place, Hilo, HI 96720, USA}
\affil[f]{Steward Observatory, University of Arizona, Tucson, AZ 87521, USA}
\affil[g]{College of Optical Sciences, University of Arizona, Tucson, AZ 87521, USA}
\affil[h]{Astrobiology Center, 2 Chome-21-1, Osawa, Mitaka, Tokyo, 181-8588, Japan}
\affil[i]{Physikalisches Institut, Universit\"at Bern, Gesellschaftsstrasse 6, CH-3012 Bern, Switzerland}

\authorinfo{Send correspondence to C.L. E-mail: christophe.lovis@unige.ch}

\begin{document} 
\maketitle

\begin{abstract}
RISTRETTO is a visible high-resolution spectrograph fed by an extreme adaptive optics (XAO) system, to be proposed as a visitor instrument on ESO VLT. The main science goal of RISTRETTO is the detection and atmospheric characterization of exoplanets in reflected light, in particular the temperate rocky planet Proxima~b. RISTRETTO will be able to measure albedos and detect atmospheric features in a number of exoplanets orbiting nearby stars for the first time. It will do so by combining a high-contrast AO system working at the diffraction limit of the telescope to a high-resolution spectrograph, via a 7-spaxel integral-field unit (IFU) feeding single-mode fibers. Further science cases for RISTRETTO include the study of accreting protoplanets such as PDS 70 b \& c through spectrally-resolved H$\alpha$ emission; and spatially-resolved studies of Solar System objects such as icy moons and the ice giants Uranus and Neptune. The project is in an advanced design phase for the spectrograph and IFU/fiber-link sub-systems, and a preliminary design phase for the AO front-end. Construction of the spectrograph and IFU/fiber-link will start at the end of 2022. RISTRETTO is a pathfinder instrument in view of similar developments at ESO ELT, in particular the SCAO-IFU mode of ELT-ANDES and the future ELT-PCS instrument.
\end{abstract}


\section{INTRODUCTION}
\label{sec:intro}  

One of the overarching goals of exoplanet science is the spectroscopic characterization of rocky planets orbiting within the habitable zone of their host stars. This includes the search for potential biosignatures in the atmospheres of these planets, opening the way to the search for life elsewhere in the Universe. In fact, what actually represents a biosignature is currently a matter of debate, and no diagnostic is known to date that would unambiguously reveal the presence of life on another planet. However, this uncertainty should by no means prevent the development of instrumentation that has the {\it capability} of probing the atmospheres and surfaces of temperate rocky exoplanets. Inhabited or not, these are extremely interesting objects in any case since they will allow us to place the Earth in the broader context of similar planets in the Universe.

From an instrumental point of view, temperate rocky exoplanets are extremely challenging to probe. Two main approaches can be pursued: for transiting planets, high-precision transmission spectra can be obtained which reveal atmospheric composition. Cumulating a large number of transits is however necessary, and transiting planets are rare statistically speaking, thus orbiting relatively distant stars. Non-transiting planets can be found around very nearby stars, but require to be angularly separated from their host star to be studied.

In the coming years, the James Webb Space Telescope is expected to become the main asset to probe small exoplanets via transmission spectroscopy. For non-transiting planets, there are long-term prospects to build dedicated space missions that will be free from the disturbances caused by Earth's atmosphere. However, in the shorter term, the focus is on ground-based high-resolution spectrographs mounted on the upcoming Extremely Large Telescopes (ELTs). These can probe small exoplanets by applying cross-correlation techniques to the many individual spectral lines in their atmospheres, while being relatively immune to telluric absorption thanks to the high spectral resolution.

The present paper focuses on RISTRETTO\cite{Lovis2019,Chazelas2020}, a pathfinder instrument that will explore for the first time a number of nearby exoplanets in reflected light. RISTRETTO will use the high-contrast, high-resolution spectroscopy technique\cite{Sparks2002,Snellen2015} to both angularly resolve the planets from their star, and be able to reach the required planet-to-star contrast to reveal the planetary signal. This technique can be thought of as a two-step approach. An eXtreme Adaptive Optics (XAO) system first removes as much stellar light as possible at the location of the planet. It is then followed by a high-resolution spectrograph which disentangles the planetary signal from the remaining (still dominant) stellar signal. This can be achieved thanks to the distinct spectral content of the planet spectrum and the relative Doppler shift between the two signals.

\begin{figure} [ht]
\begin{center}
\includegraphics[width=\textwidth]{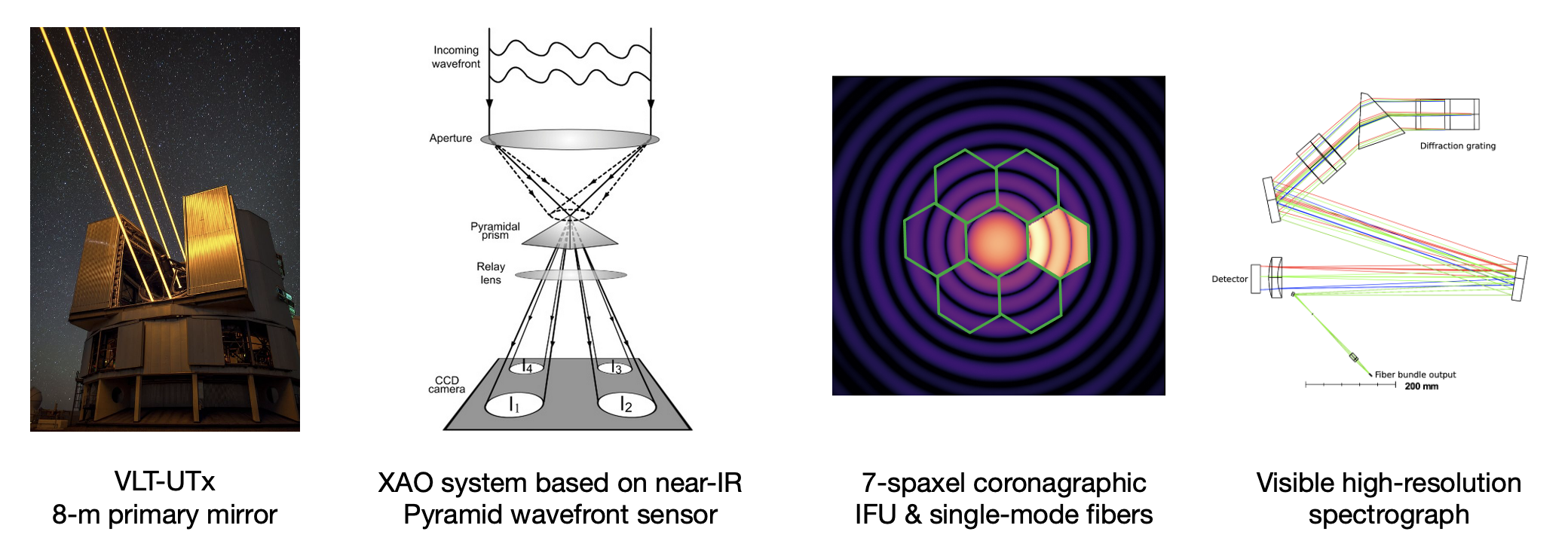}
\end{center}
\caption[example] 
{ \label{fig:schematic} 
Schematic of the RISTRETTO instrument.}
\end{figure} 

RISTRETTO is designed as a visitor instrument for ESO VLT. The instrument is split into two main parts: the back-end comprising a visible high-resolution spectrograph, the fiber link and integral-field unit (IFU); and the front-end comprising the XAO system and the coronagraph/apodizer. Fig.~\ref{fig:schematic} provides a schematic of the entire system. The spectrograph part is currently in an advanced design phase and its construction is expected to start at the end of 2022. It will then be tested on the Swiss 1.2-m telescope at La Silla Observatory (Chile), which is equipped with the KalAO adaptive optics system\cite{Hagelberg2020}. The front-end part is currently in a preliminary design phase. The target date for the installation of the full system at the VLT is 2025.

This paper describes the main science cases driving the top-level requirements for RISTRETTO. It is accompanied by three other papers covering the technical aspects of the instrument, namely the XAO system and coronagraph design (Blind et al., paper 12185-269), the coronagraph and IFU prototyping (K\"uhn et al., paper 12188-64), and the spectrograph design (Chazelas et al., paper 12184-181).

\section{EXOPLANETS IN REFLECTED LIGHT}
\label{sec:exoplanets}

The reflected-light contrast $C$ of an exoplanet with respect to its host star can be expressed as

\begin{equation}
\label{eq:contrast}
C = \left(\frac{R_p}{a}\right)^2 A_g(\lambda) \, g(\alpha) \, ,
\end{equation}
where $R_p$ is the planet radius, $a$ is the semi-major axis of the planetary orbit, $A_g(\lambda)$ is the wavelength-dependent geometric albedo, and $g(\alpha)$ is the phase function (see Ref.~\citenum{Lovis2017}).

\begin{figure} [ht]
\begin{center}
\includegraphics[width=\textwidth]{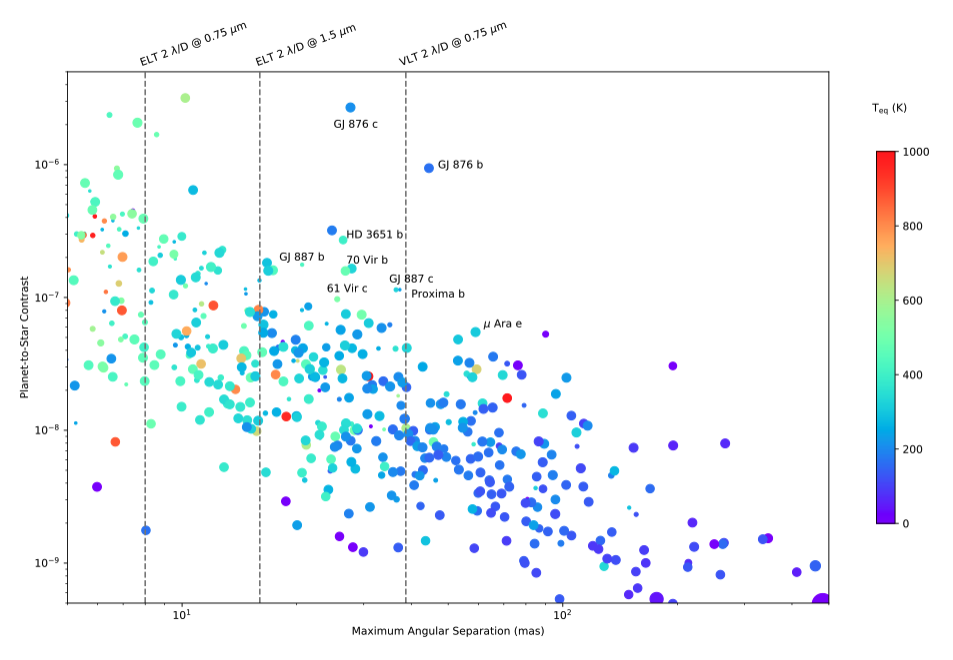}
\end{center}
\caption[example] 
{ \label{fig:contrast} 
Expected reflected-light contrast of known exoplanets as a function of the maximum angular separation from their host star. Dot size increases with planet mass, while the color scale indicates equilibrium temperature. Dashed vertical lines show the 2 $\lambda / D$ threshold for the VLT in $I$-band and the ELT in $I$- and $H$-bands. A number of exoplanets with favorable angular separation and contrast are labelled. Notably, the temperate and rocky Proxima b is accessible to the VLT in $I$-band, at an expected contrast of $10^{-7}$.}
\end{figure}

We compute this quantity for all known exoplanets observable from Paranal ($\delta$ $<$ $+$25 deg), assuming an average geometric albedo of 0.4 and observations at quadrature considering a Lambertian reflection law ($g(\alpha)$=1/$\pi$). The results are shown in Fig.~\ref{fig:contrast}. They are plotted as a function of the maximum angular separation from the host star $\theta_{max}=a/d$ (i.e., at quadrature), where $d$ is the distance to the system. Fig.~\ref{fig:contrast} reveals a number of exoplanets with potentially favorable separations and contrasts. As an example, GJ 876 b \& c are two temperate giant planets orbiting a nearby M dwarf that stand out as particularly attractive targets. Moreover, the closest exoplanet to us, the temperate and rocky Proxima~b, is by far the most accessible small planet, at a maximum separation of 37 mas and an expected contrast of 10$^{-7}$.

The existence of Proxima~b and the fact that it is just resolvable by the VLT in $I$-band is at the origin of the RISTRETTO idea. Detailed simulations showed that Proxima~b could indeed be accessible with the VLT using an optimized instrument combining a state-of-the-art XAO system and a visible high-resolution spectrograph\cite{Lovis2017}. While Proxima~b is chosen as the defining science case for RISTRETTO, this specific application has to be understood in a broader context:

\begin{itemize}
\item Before attempting Proxima~b, RISTRETTO will be able to detect and characterize a number of "easier" exoplanets for the first time. It will thus pioneer reflected-light exoplanet spectroscopy and explore a diverse sample of objects ranging from cold Jupiter-like planets to warm super-Earths (see below).
\item RISTRETTO paves the way for the high-contrast, high-resolution technique to be deployed on the upcoming ELTs, which will rely on the same approach to probe habitable exoplanets and search for signs of life. This applies in particular to the ANDES\cite{Marconi2021} and PCS\cite{Kasper2021} instruments on the European ELT. 
\end{itemize}

Simulations probing the detectability of Proxima~b with RISTRETTO at the VLT yield the following top-level requirements for the instrument:

\begin{itemize}
\item Spectrograph wavelength range: 620-840 nm
\item Spectral resolution $R$ $>$ 140,000
\item Inner working angle of the system $<=$ 2 $\lambda/D$
\item Total system throughput $>$ 5\%
\item Planet coupling efficiency into the off-axis spaxels $>$ 50\%
\item Stellar coupling into the off-axis spaxels $<$ 10$^{-4}$ (instantaneous, not post-processed)
\end{itemize}

Based on these requirements, we derive the estimated number of VLT nights required to detect the reflected-light spectrum of all exoplanets shown in Fig.~\ref{fig:contrast}. Here we define detection as reaching SNR=5 on the planet cross-correlation function (CCF) using 1000 individual spectral lines. Note that these lines may come from the reflected stellar spectrum or from a given molecule of interest in the exoplanet atmosphere. In the first case the detection will constrain the average albedo of the planet over the chosen bandpass as well as the planet true mass (as opposed to minimum mass). In the second case, the detection will directly constrain the atmospheric composition and provide information on pressure and temperature.

\begin{figure} [ht]
\begin{center}
\includegraphics[width=\textwidth]{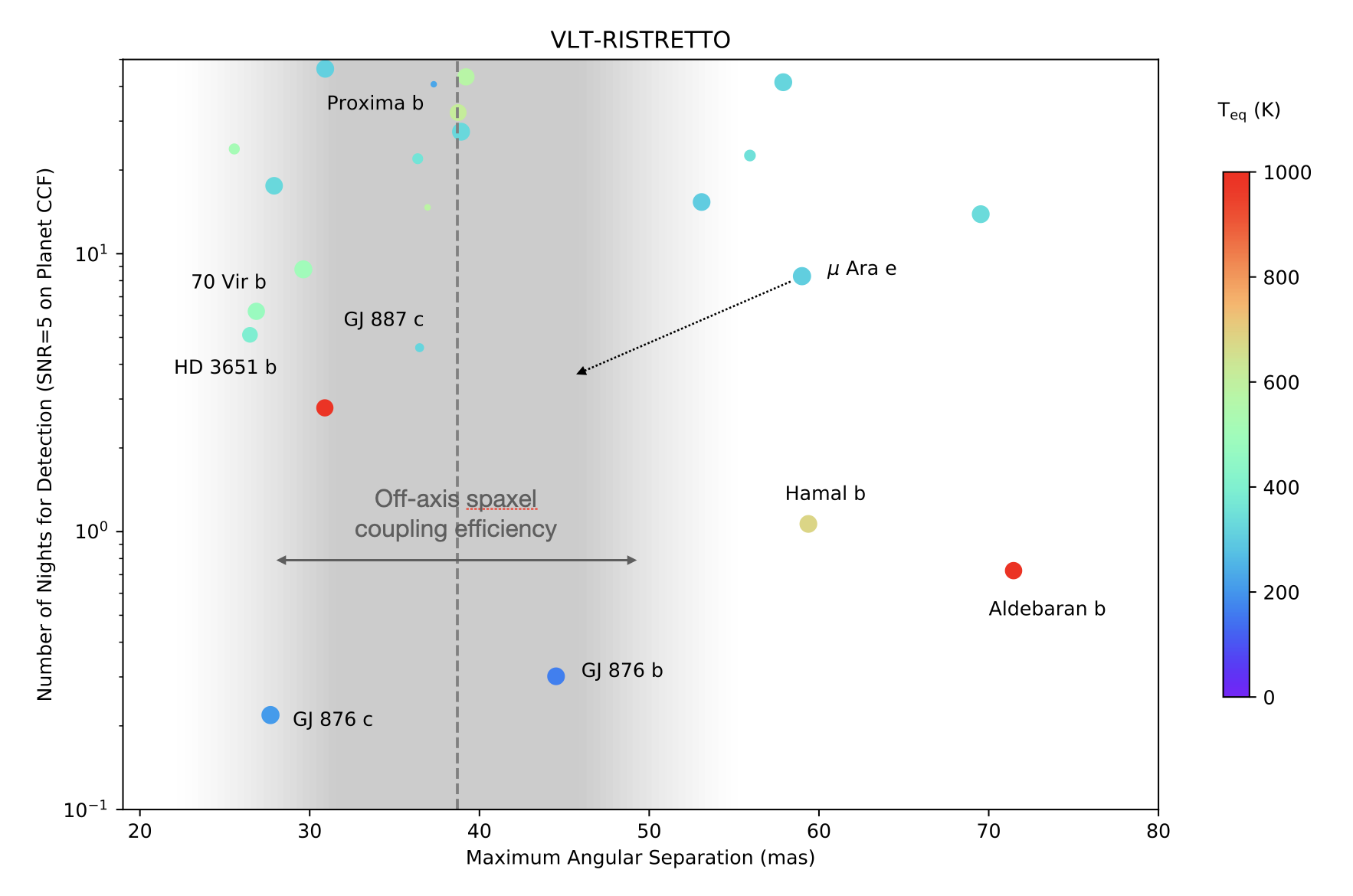}
\end{center}
\caption[example] 
{ \label{fig:target_list} 
Current RISTRETTO target list for the characterization of exoplanets in reflected light. The plot shows the estimated number of VLT nights required to achieve SNR=5 on the planet cross-correlation function, as a function of the maximum angular separation from the host star. The dashed vertical line indicates the center of the off-axis spaxels, while the shaded area represents the planet coupling efficiency into the off-axis spaxels. The observing time computation takes into account the degraded coupling efficiency at separations smaller than the spaxel center. Planets at larger maximum separations can be observed at orbital phases closer to superior conjunction, when they are located near spaxel centers and have a higher reflected-light contrast (fuller illumination). This is illustrated by the arrow on planet $\mu$ Ara e.}
\end{figure} 

Results are shown in Fig.~\ref{fig:target_list}. We can see that the gas giants GJ 876 b \& c are by far the easiest objects to characterize, as they can be detected in 2-3 hours of observing time. Interestingly, the giant planet candidates orbiting the red giants Aldebaran and Hamal could also be detected in less than a night when observed closer to superior conjunction, i.e. when they fall within the off-axis spaxels of RISTRETTO. A number of other giant planets can be detected in a few nights of observations, such as $\mu$ Ara e, 70 Vir b and HD 3651 b. Going towards smaller planets, the warm super-Earth GJ 887 c ($m \sin{i}$ = 7.6 $M_{\oplus}$) could be detected in about 5 nights of integration. Finally, Proxima~b can be detected in about 40 nights of VLT time. Although it may sound like a lot of telescope time for a single object, one should remember that the characterization of potentially habitable exoplanets is one of the top science drivers for several major space- and ground-based observatories, among which JWST and the European ELT. These facilities will invest a significant amount of time on such objects. In this context, we note that 40 VLT nights represent about 2 ELT nights in terms of photon collecting power. Moreover, the current baseline design of RISTRETTO shows performances that exceed the requirements listed above, potentially enabling a gain of a factor $\sim$2 in exposure time (see Blind et al., paper 12185-269).

In summary, RISTRETTO will be able to detect and characterize a sample of at least 10 exoplanets orbiting very nearby stars. These planets represent a diverse sample in terms of mass (from gas giants to rocky worlds), equilibrium temperature (from hot to cold), orbital distance and host star type. RISTRETTO will thus for the first time demonstrate the feasibility of reflected-light exoplanet spectroscopy, paving the way for future instruments such as ANDES and PCS on the European ELT.

\section{ACCRETING PROTOPLANETS IN H-ALPHA}
\label{sec:protoplanets}

Beyond exoplanets in reflected light, we are currently exploring additional science cases to exploit the full potential of RISTRETTO. A promising science case is the detection and characterization of accreting protoplanets through spectrally-resolved H$\alpha$ observations. Accretion processes onto forming protoplanets indeed trigger powerful H$\alpha$ emission which can be directly detected if it can be disentangled from the stellar H$\alpha$ emission. This has been achieved for the PDS 70 system using the MUSE integral-field spectrograph at the VLT\cite{Haffert2019}. Two accreting protoplanets are directly detected in H$\alpha$ around the young star PDS 70, at angular separations of about 180 and 230 mas. In principle, a lot of useful information on planet formation processes could be obtained from detailed studies of planetary H$\alpha$ line profiles. At high spectral resolution, velocity fields within the accretion flows could be directly measured through their Doppler shifts, and the geometry of the flows could be modelled in some details together with the mass accretion rates\cite{Marleau2022,Aoyama2021}. More generally, one can also use the H$\alpha$ line to detect yet unknown exoplanets around a population of young stars, especially at small inner working angles where standard high-contrast imaging instruments may have limited sensitivity.

\begin{figure} [ht]
\begin{center}
\includegraphics[width=15cm]{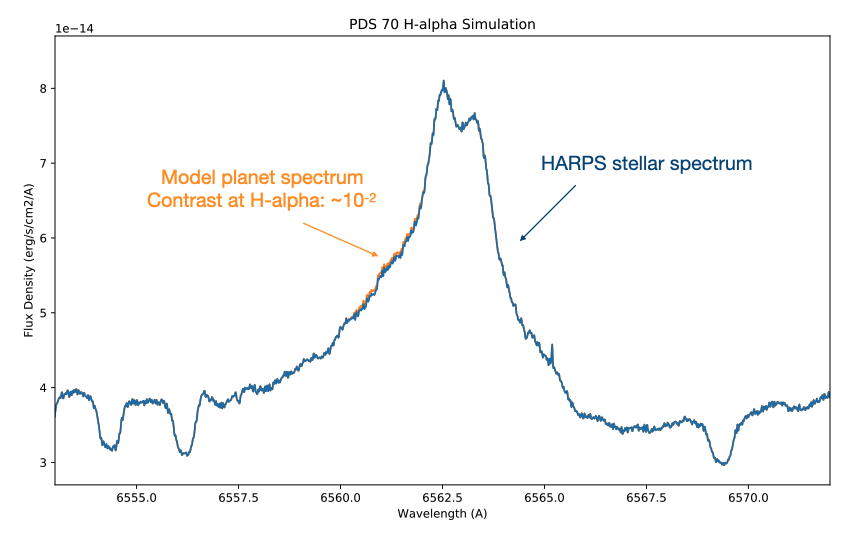}
\end{center}
\caption[example] 
{ \label{fig:PDS70_1} 
Simulated RISTRETTO observation of the young star PDS 70. A real high-resolution HARPS spectrum of PDS 70 is used as stellar spectrum in the H$\alpha$ region. A planetary accretion H$\alpha$ emission spectrum matching the measured properties of PDS 70 b is added to the simulation (from Y. Aoyama, private communication). The resulting planet-to-star contrast at H$\alpha$ is about 10$^{-2}$.}
\end{figure} 

To explore these possibilities, we developed a simple simulation of PDS 70 observed with RISTRETTO. We first retrieve from the ESO archive a number of actual high-resolution HARPS spectra of PDS 70. We then build a properly flux-calibrated master spectrum of the stellar H$\alpha$ line, which is shown in Fig.~\ref{fig:PDS70_1}. We add to it a model high-resolution emission spectrum from an accreting planet that matches the measured properties of PDS 70 b (from Y. Aoyama, private communication). Interestingly, the planet spectrum can be seen by eye in Fig.~\ref{fig:PDS70_1} because the planet-to-star contrast at H$\alpha $ is as high as 10$^{-2}$, i.e. orders of magnitude larger than the typical contrasts of young planets observed in thermal emission.

\begin{figure} [ht]
\begin{center}
\includegraphics[width=15cm]{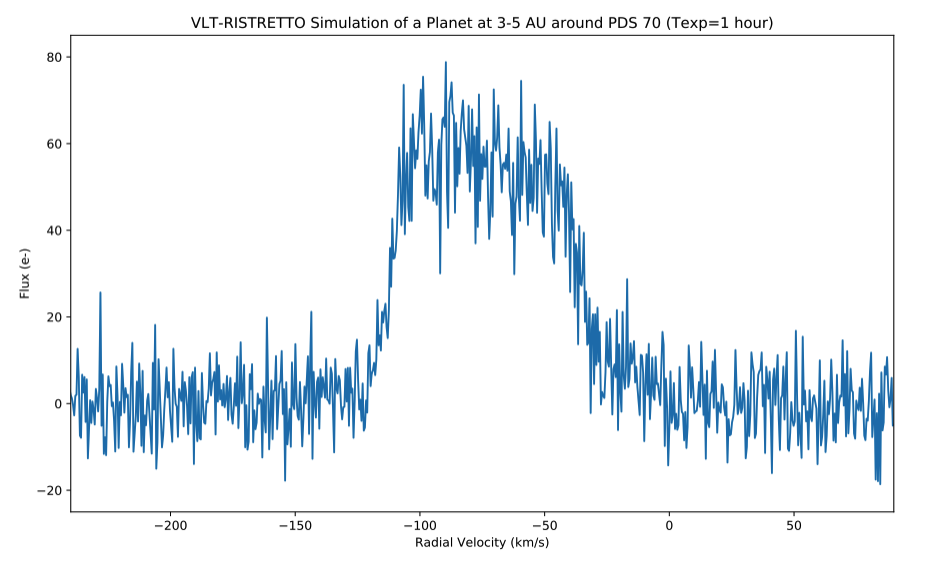}
\end{center}
\caption[example] 
{ \label{fig:PDS70_2} 
Simulated RISTRETTO H$\alpha$ observation of an accreting protoplanet orbiting PDS 70, with similar properties as PDS 70 b, but located at only 3-5 AU from the star (i.e. within the off-axis spaxels). The spectrally-resolved H$\alpha$ line profile of the accretion flow onto the planet is clearly detected in 1 hour of exposure time.}
\end{figure} 

We then proceed with adding photon and detector noise to the simulated data assuming realistic performances for RISTRETTO, considering in particular the lower performance of the AO system on a faint star such as PDS 70. We finally obtain the planet-only spectrum by subtracting a high-SNR star-only spectrum from an observed star$+$planet spectrum in an off-axis spaxel of RISTRETTO. This simulates a protoplanet that would have similar properties as PDS 70 b, but that would be located much closer to the star, on a 3-5 AU orbit. The result is shown in Fig.~\ref{fig:PDS70_2}. The spectrally-resolved H$\alpha$ line profile of the accretion flow onto the planet is clearly detected in 1 hour of exposure time. We conclude that RISTRETTO could not only characterize already-known protoplanets in H$\alpha$, but could also search for new accreting planets around young stars. It could do so at orbital separations corresponding to a few AUs at the distance of typical star-forming regions, i.e. at orbital distances where planet formation models expect that most giant planets form. In summary, RISTRETTO could usefully probe the close stellar environment (30-50 mas) around a sample of young stars, potentially revealing new protoplanets that are beyond the reach of current state-of-the-art direct imaging facilities.

\section{SOLAR SYSTEM SCIENCE}
\label{sec:solar_system}

\begin{figure} [ht]
\begin{center}
\includegraphics[width=\textwidth]{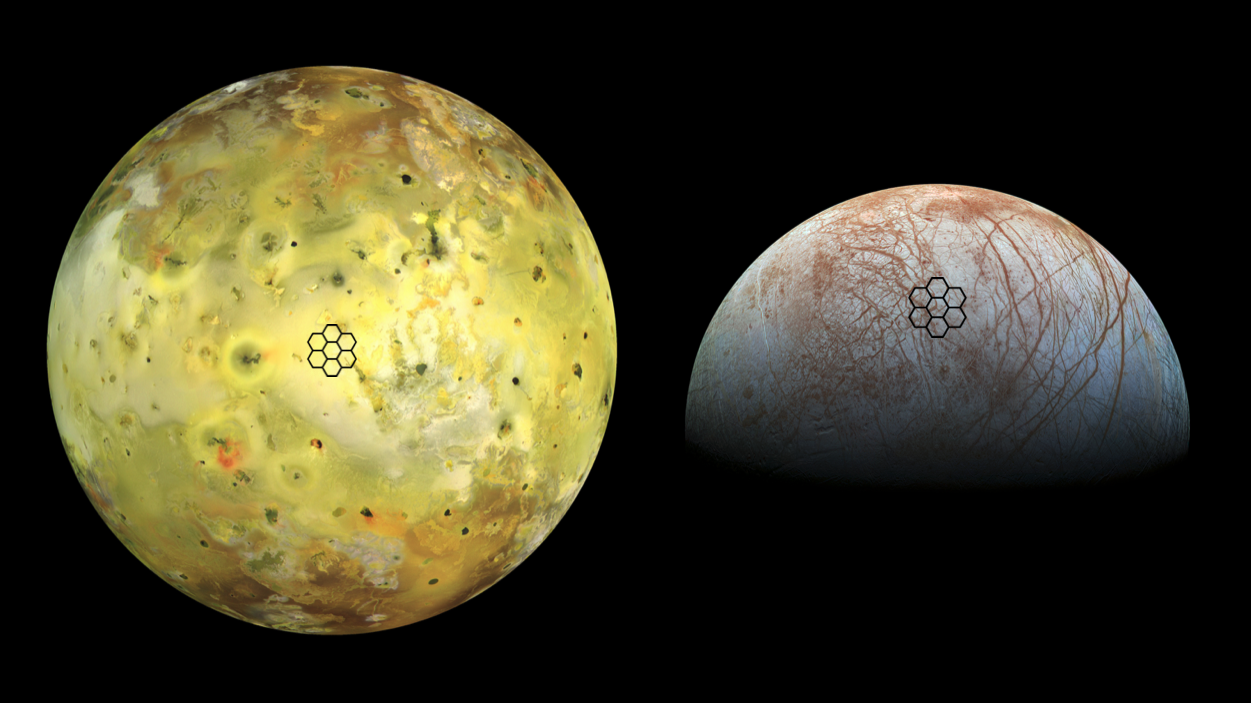}
\end{center}
\caption[example] 
{ \label{fig:Io_Europa} 
Projection of the RISTRETTO IFU onto the Jovian moons Io (left) and Europa (right). The angular size of the two moons as seen from Earth is 1.2 arcsec and 1.0 arcsec, respectively. The spatial resolution achieved by RISTRETTO makes it possible to study individual surface features on these icy moons.}
\end{figure}

We are also exploring whether RISTRETTO could produce meaningful observations of Solar System objects, in particular the icy moons of Jupiter and Saturn, as well as the ice giants Uranus and Neptune. In this context, the uniqueness of RISTRETTO lies in the high spatial resolution ($\sim$40 mas) it can offer from the ground, coupled to the high spectral resolution. We illustrate the achievable spatial resolution by projecting the RISTRETTO IFU onto the Jovian moons Io and Europa in Fig.~\ref{fig:Io_Europa}. As can be seen, it would be possible to study individual surface features on these moons, such as volcanoes on Io. It remains to be explored whether there exist interesting spectral features in the RISTRETTO wavelength range (620-840 nm) which could be used to complement the current knowledge we have of these moons.

Another potential application of RISTRETTO on Solar System objects is Doppler velocimetry to probe local atmospheric phenomena such as wind patterns and planetary oscillations. Indeed RISTRETTO should be able to measure radial velocities on the reflected solar spectrum to a precision of a few m s$^{-1}$. This could be particularly interesting for Uranus and Neptune, which are still poorly characterized.

\section{CONCLUSION}
\label{sec:conclusion}

RISTRETTO is an innovative instrument that will provide high-resolution spectroscopy at the diffraction limit of the telescope for the first time. It will pioneer reflected-light spectroscopy of exoplanets, including Proxima~b. In this respect it is a pathfinder towards the ELT instruments ANDES and PCS. RISTRETTO will also address a number of other science cases in an original way, in particular accreting protoplanets in H$\alpha$ and Solar System observations.

We refer the reader to the accompanying papers 12185-269 (Blind et al.), 12188-64 (K\"uhn et al.), and 12184-181 (Chazelas et al.) for details on the design of the different RISTRETTO sub-systems.

\acknowledgments 

This work has been carried out within the framework of the National Centre of Competence in Research PlanetS supported by the Swiss National Science Foundation under grants 51NF40\_182901 and 51NF40\_205606. The RISTRETTO project was partially funded through the SNSF FLARE programme for large infrastructures under grants 20FL21\_173604 and 20FL20\_186177. The authors acknowledge the financial support of the SNSF.

\bibliography{report} 

\begin{thebibliography}{10}

\bibitem{Lovis2019}
{Lovis}, C., ``{Reflected-light spectroscopy of nearby exoplanets with
  RISTRETTO at the VLT},'' in [{\em The Very Large Telescope in
  2030}{\nolinebreak\hspace{0.1em}]},   41 (July 2019).

\bibitem{Chazelas2020}
{Chazelas}, B., {Lovis}, C., {Blind}, N., {K{\"u}hn}, J., {Genolet}, L.,
  {Hughes}, I., {Turbet}, M., {Hagelberg}, J., {Restori}, N., {Kasper}, M., and
  {Cerpa Urra}, N.~N., ``{RISTRETTO: a pathfinder instrument for exoplanet
  atmosphere characterization},'' in [{\em Society of Photo-Optical
  Instrumentation Engineers (SPIE) Conference
  Series}{\nolinebreak\hspace{0.1em}]},  {\em Society of Photo-Optical
  Instrumentation Engineers (SPIE) Conference Series} {\bf 11448},  1144875
  (Dec. 2020).

\bibitem{Sparks2002}
{Sparks}, W.~B. and {Ford}, H.~C., ``{Imaging Spectroscopy for Extrasolar
  Planet Detection},'' {\em The Astrophysical Journal}~{\bf 578},  543--564
  (Oct. 2002).

\bibitem{Snellen2015}
{Snellen}, I., {de Kok}, R., {Birkby}, J.~L., {Brandl}, B., {Brogi}, M.,
  {Keller}, C., {Kenworthy}, M., {Schwarz}, H., and {Stuik}, R., ``{Combining
  high-dispersion spectroscopy with high contrast imaging: Probing rocky
  planets around our nearest neighbors},'' {\em Astronomy \& Astrophysics}~{\bf
  576},  A59 (Apr. 2015).

\bibitem{Hagelberg2020}
{Hagelberg}, J., {Restori}, N., {Wildi}, F., {Chazelas}, B., {Baranec}, C.,
  {Guyon}, O., {Genolet}, L., {Sordet}, M., and {Riddle}, R., ``{KalAO the
  swift adaptive optics imager on the 1.2m Euler Swiss telescope in La Silla,
  Chile},'' in [{\em Society of Photo-Optical Instrumentation Engineers (SPIE)
  Conference Series}{\nolinebreak\hspace{0.1em}]},  {\em Society of
  Photo-Optical Instrumentation Engineers (SPIE) Conference Series} {\bf
  11448},  114487G (Dec. 2020).

\bibitem{Lovis2017}
{Lovis}, C., {Snellen}, I., {Mouillet}, D., {Pepe}, F., {Wildi}, F.,
  {Astudillo-Defru}, N., {Beuzit}, J.~L., {Bonfils}, X., {Cheetham}, A.,
  {Conod}, U., {Delfosse}, X., {Ehrenreich}, D., {Figueira}, P., {Forveille},
  T., {Martins}, J.~H.~C., {Quanz}, S.~P., {Santos}, N.~C., {Schmid}, H.~M.,
  {S{\'e}gransan}, D., and {Udry}, S., ``{Atmospheric characterization of
  Proxima b by coupling the SPHERE high-contrast imager to the ESPRESSO
  spectrograph},'' {\em Astronomy \& Astrophysics}~{\bf 599},  A16 (Mar. 2017).

\bibitem{Marconi2021}
{Marconi}, A., {Abreu}, M., {Adibekyan}, V., {Aliverti}, M., {Allende Prieto},
  C., {Amado}, P., {Amate}, M., {Artigau}, E., {Augusto}, S., {Barros}, S.,
  {Becerril}, S., {Benneke}, B., {Bergin}, E., {Berio}, P., {Bezawada}, N.,
  {Boisse}, I., {Bonfils}, X., {Bouchy}, F., {Broeg}, C., {Cabral}, A.,
  {Calvo-Ortega}, R., {Canto Martins}, B.~L., {Chazelas}, B., {Chiavassa}, A.,
  {Christensen}, L., {Cirami}, R., {Coretti}, I., {Covino}, S., {Cresci}, G.,
  {Cristiani}, S., {Cunha Parro}, V., {Cupani}, G., {de Castro Le{\~a}o}, I.,
  {Renan de Medeiros}, J., {Furlande Souza}, M.~A., {Di Marcantonio}, P., {Di
  Varano}, I., {D'Odorico}, V., {Doyon}, R., {Drass}, H., {Figueira}, P.,
  {Belen Fragoso}, A., {Uldall Fynbo}, J.~P., {Gallo}, E., {Genoni}, M.,
  {Gonz{\'a}lez Hern{\'a}ndez}, J., {Haehnelt}, M., {Hlavacek-Larrondo}, J.,
  {Hughes}, I., {Huke}, P., {Humphrey}, A., {Kjeldsen}, H., {Korn}, A.,
  {Kouach}, D., {Landoni}, M., {Liske}, J., {Lovis}, C., {Lunney}, D.,
  {Maiolino}, R., {Malo}, L., {Marquart}, T., {Martins}, C., {Mason}, E.,
  {Molaro}, P., {Monnier}, J., {Monteiro}, M., {Mordasini}, C., {Morris}, T.,
  {Mucciarelli}, A., {Murray}, G., {Niedzielski}, A., {Nunes}, N., {Oliva}, E.,
  {Origlia}, L., {Pall{\'e}}, E., {Pariani}, G., {Parr-Burman}, P.,
  {Pe{\~n}ate}, J., {Pepe}, F., {Pinna}, E., {Piskunov}, N., {Rasilla
  Pi{\~n}eiro}, J.~L., {Rebolo}, R., {Rees}, P., {Reiners}, A., {Riva}, M.,
  {Romano}, D., {Rousseau}, S., {Sanna}, N., {Santos}, N., {Sarajlic}, M.,
  {Shen}, T.~C., {Sortino}, F., {Sosnowska}, D., {Sousa}, S., {Stempels}, E.,
  {Strassmeier}, K., {Tenegi}, F., {Tozzi}, A., {Udry}, S., {Valenziano}, L.,
  {Vanzi}, L., {Weber}, M., {Woche}, M., {Xompero}, M., {Zackrisson}, E., and
  {Zapatero Osorio}, M.~R., ``{HIRES, the High-resolution Spectrograph for the
  ELT},'' {\em The Messenger}~{\bf 182},  27--32 (Mar. 2021).

\bibitem{Kasper2021}
{Kasper}, M., {Cerpa Urra}, N., {Pathak}, P., {Bonse}, M., {Nousiainen}, J.,
  {Engler}, B., {Heritier}, C.~T., {Kammerer}, J., {Leveratto}, S., {Rajani},
  C., {Bristow}, P., {Le Louarn}, M., {Madec}, P.~Y., {Str{\"o}bele}, S.,
  {Verinaud}, C., {Glauser}, A., {Quanz}, S.~P., {Helin}, T., {Keller}, C.,
  {Snik}, F., {Boccaletti}, A., {Chauvin}, G., {Mouillet}, D., {Kulcs{\'a}r},
  C., and {Raynaud}, H.~F., ``{PCS {\textemdash} A Roadmap for Exoearth Imaging
  with the ELT},'' {\em The Messenger}~{\bf 182},  38--43 (Mar. 2021).

\bibitem{Haffert2019}
{Haffert}, S.~Y., {Bohn}, A.~J., {de Boer}, J., {Snellen}, I.~A.~G.,
  {Brinchmann}, J., {Girard}, J.~H., {Keller}, C.~U., and {Bacon}, R., ``{Two
  accreting protoplanets around the young star PDS 70},'' {\em Nature
  Astronomy}~{\bf 3},  749--754 (June 2019).

\bibitem{Marleau2022}
{Marleau}, G.~D., {Aoyama}, Y., {Kuiper}, R., {Follette}, K., {Turner}, N.~J.,
  {Cugno}, G., {Manara}, C.~F., {Haffert}, S.~Y., {Kitzmann}, D., {Ringqvist},
  S.~C., {Wagner}, K.~R., {van Boekel}, R., {Sallum}, S., {Janson}, M.,
  {Schmidt}, T.~O.~B., {Venuti}, L., {Lovis}, C., and {Mordasini}, C.,
  ``{Accreting protoplanets: Spectral signatures and magnitude of gas and dust
  extinction at H {\ensuremath{\alpha}}},'' {\em Astronomy \&
  Astrophysics}~{\bf 657},  A38 (Jan. 2022).

\bibitem{Aoyama2021}
{Aoyama}, Y., {Marleau}, G.-D., {Ikoma}, M., and {Mordasini}, C., ``{Comparison
  of Planetary H{\ensuremath{\alpha}}-emission Models: A New Correlation with
  Accretion Luminosity},'' {\em The Astrophysical Journal Letters}~{\bf 917},
  L30 (Aug. 2021).

\end{thebibliography}
\bibliographystyle{spiebib} 

\end{document}